\documentclass[12pt]{article}
\usepackage{epsfig}
\usepackage{color}
\def\be{\begin{equation}}
\def\ee{\end{equation}}
\def\bea{\begin{eqnarray}}
\def\eea{\end{eqnarray}}
\usepackage{graphicx}
\usepackage{amsmath,amssymb}

\catcode`\@=11
\def\lsim{\mathrel{\mathpalette\@versim<}}
\def\gsim{\mathrel{\mathpalette\@versim>}}
\def\@versim#1#2{\vcenter{\offinterlineskip
\ialign{$\m@th#1\hfil##\hfil$\crcr#2\crcr\sim\crcr } }}
\catcode`\@=12

\catcode`@=12
\begin{document}
\thispagestyle{empty}
\begin{flushright}
UCRHEP-T560\\
December 2015\
\end{flushright}
\vspace{0.3in}
\begin{center}
{\Large \bf New U(1) Gauge Model of Radiative Lepton Masses\\ 
with Sterile Neutrino and Dark Matter\\}
\vspace{0.5in}
{\bf Rathin Adhikari\\}
\vspace{0.1in}
{\sl Centre for Theoretical Physics, Jamia Millia Islamia (Central 
University),\\ Jamia Nagar, New Delhi 110025, India\\}
\vspace{0.2in}
{\bf Debasish Borah\\}
\vspace{0.1in}
{\sl Department of Physics, Indian Institute of Technology Guwahati, Assam 781039, India\\}
\vspace{0.2in}
{\bf Ernest Ma\\}
\vspace{0.1in}
{\sl Physics \& Astronomy Department and Graduate Division,\\ 
University of California, Riverside, California 92521, USA\\}
\vspace{0.1in}
{\sl HKUST Jockey Club Institute for Advanced Study,\\
Hong Kong University of Science and Technology, Hong Kong, China\\}
\end{center}
\vspace{0.5in}

\begin{abstract}\
An anomaly-free U(1) gauge extension of the standard model (SM) 
is presented. Only one Higgs doublet with a nonzero vacuum expectation is 
required as in the SM.  New fermions and scalars as well as all SM particles 
transform nontrivially under this U(1), resulting in a model of three 
active neutrinos and one sterile neutrino, all acquiring radiative masses. 
Charged-lepton masses are also radiative as well as the mixing between 
active and sterile neutrinos.  At the same time, a residual $Z_2$ symmetry 
of the U(1) gauge symmetry remains exact, allowing for the existence of 
dark matter. 
\end{abstract}

\newpage
\baselineskip 24pt

\begin{figure}[htb]
\vspace*{-3cm}
\hspace*{-3cm}
\includegraphics[scale=1.0]{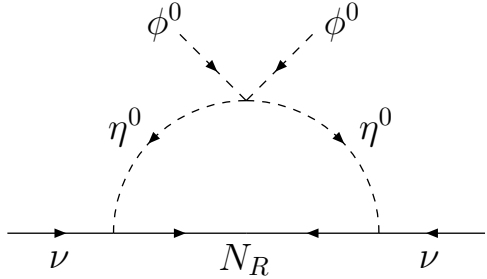}
\vspace*{-21.0cm}
\caption{One-loop ``scotogenic'' neutrino mass.}
\label{fig1}
\end{figure}

The notion that neutrino mass is connected to dark matter has motivated a 
large number of studies in recent years.  The simplest realization is the 
one-loop ``scotogenic'' model~\cite{m06}, where the standard model (SM) of 
quarks and leptons is augmented with a second scalar doublet $(\eta^+,\eta^0)$ 
and three neutral singlet fermions $N_R$ as shown in Fig.~\ref{fig1}.
Under an exactly conserved discrete $Z_2$ symmetry, $(\eta^+,\eta^0)$ and 
$N$ are odd, allowing thus the existence of dark matter (DM).  Whereas such 
models are viable phenomenologically, a deeper theoretical understanding 
of the origin of this connection is clearly desirable.

Another important input to this framework is the 2012 discovery of the 125 
GeV particle~\cite{atlas12,cms12} at the Large Hadron Collider (LHC) which 
looks very much like the one Higgs boson of the SM.  This means that any 
extension of the SM should aim for a natural explanation of why electroweak 
symmetry breaking appears to be embodied completely in one Higgs scalar 
doublet and no more.

To these ends, we propose in this paper an anomaly-free $U(1)_X$ gauge 
extension of the SM with three active and one sterile neutrinos. 
Whereas there exist many studies on light sterile neutrino 
masses~\cite{sterile,ba14}, we consider here for the first time 
the case where all masses and mixing of active and sterile neutrinos 
are generated in one loop through dark 
matter, which is stabilized by a residual $Z_2$ symmetry of the spontaneously 
broken $U(1)_X$ gauge symmetry.  To maintain the hypothesis of only one 
electroweak symmetry breaking Higgs doublet (which couples directly only 
to quarks in this model), charged-lepton masses are also radiatively generated 
through dark matter.

The $U(1)_X$ gauge symmetry being considered is a variation of Model (C) of 
Ref.~\cite{aem09}.  It has its origin from the  
observation~\cite{bbb86,m02,mr02,bd05} that replacing the neutral singlet 
fermion $N$ of the Type I seesaw for neutrino mass with the fermion triplet 
$(\Sigma^+,\Sigma^0,\Sigma^-)$ of the Type III seesaw also results in a 
possible U(1) gauge extension.  The former is the well-known $B-L$, the 
latter is the model of Ref.~\cite{m02}, where there is one $\Sigma$ for 
each of the three families of quarks and leptons.  Here we consider a total 
of only two $\Sigma$'s, in which case several $N$'s of different $U(1)_X$ 
charges must be added to render the model anomaly-free.  Model (C) of 
Ref.~\cite{aem09} is the first such example with three $N$'s.  It allows 
radiative neutrino masses with dark matter~\cite{ba12,bd15,bda15}.  It may 
also accommodate a sterile neutrino with radiative mass~\cite{ba14}, but then 
dark matter is lost.  Here we choose to satisfy the anomaly-free conditions 
with three different $N$'s.  In so doing, we obtain a model with dark matter 
as well as radiative masses and mixing for three active and one sterile 
neutrinos as described below.

Under $U(1)_X$, let three families of 
$(u,d)_L$, $u_R$, $d_R$, $(\nu,e)_L$, $e_R$ transform as $n_{1,2,3,4,5}$ 
respectively.  We add two copies of $(\Sigma^+,\Sigma^0,\Sigma^-)$, each 
transforming as $n_6$.  As shown in Ref.~\cite{aem09}, the conditions 
for the absence of axial-vector anomalies in the presence of $U(1)_X$ 
determine $n_{2,3,5,6}$ in terms of $n_1$ and $n_4$ with $3n_1 + n_4 \neq 0$. 
To satisfy the $\sum U(1)_X^3 = 0$ condition and the $\sum U(1)_X = 0$ 
condition due to the mixed gravitational-gauge anomaly, three neutral singlet 
$N$'s are added.  In Model (C) of Ref.~\cite{aem09}, their charges [in units 
of $(3n_1+n_4)/8$] are $(3,2,-5)$.  Here we choose instead $(-6,1,5)$.  
Note that 
\begin{eqnarray}
&& 3^3 + 2^3 + (-5)^3 = -90, ~~~ 3+2-5 = 0, \\ 
&& (-6)^3 + 1^3 + 5^3 = -90, ~~~ -6+1+5 = 0,
\end{eqnarray}
i.e. they give identical contributions to the anomaly-free conditions.
However, the latter choice leads to the new model with particle content 
given in Table~1.
\begin{table}[htb]
\begin{center}
\begin{tabular}{|c|c|c|c|}
\hline
particle & $a_1$ & $a_4$ & $Z_2$ \\ 
\hline \hline
$(u,d)_L$ & 1 & 0 & + \\ 
$u_R$ & 7/4 & --3/4 & + \\ 
$d_R$ & 1/4 & 3/4 & + \\ 
\hline
$(\nu,l)_L$ & 0 & 1 & + \\ 
$l_R$ & -9/4 & 5/4 & + \\ 
\hline
$\Sigma^{(+,0,-)}_{1R,2R}$ & 9/8 & 3/8 & -- \\ 
\hline
$N_{R}$ & -9/4 & -3/4 & + \\ 
\hline 
$S_{1R}$ & 3/8 & 1/8 & --  \\ 
$S_{2R}$ & 15/8 & 5/8 & -- \\ 
\hline \hline
$\phi^{(+,0)}$ & 3/4 & --3/4 & + \\ 
\hline
$\eta^{(+,0)}_1$ & 3/8 & --7/8 & -- \\  
$\eta^{(+,0)}_2$ & 9/8 & --5/8 & -- \\ 
\hline
$\chi^0_1$ & 3/4 & 1/4 & + \\
$\chi^0_2$ & 9/4 & 3/4 & + \\
\hline
$\chi^0_3$ & 3/8 & 1/8 & -- \\ 
$\chi_4^+$ & 3/8 & --15/8 & -- \\ 
$\xi^{(++,+,0)}$ & 9/8 & --13/8 & -- \\
\hline 
\label{table1}
\end{tabular}
\end{center}
\caption{Particle content of proposed model with $U(1)_X$ assignment given by
$a_1 n_1 + a_4 n_4$ where $3n_1 + n_4 \neq 0$.}
\end{table}
The various scalars have been added to allow for all fermions to acquire 
nonzero masses.  An automatic residual $Z_2$ symmetry is obtained as 
$U(1)_X$ is spontaneously broken by $\chi^0_{1,2}$.  The three neutral 
singlet fermions are relabelled $N_R$ and $S_{1R,2R}$.

\begin{figure}[htb]
\vspace*{-2cm}
\hspace*{-3cm}
\includegraphics[scale=1.0]{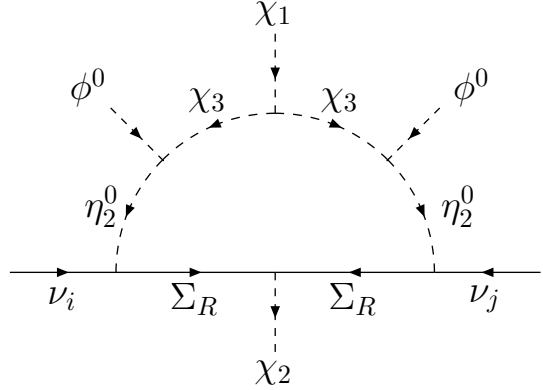}
\vspace*{-21.0cm}
\caption{One-loop active neutrino mass from $\Sigma$.}
\label{fig2}
\end{figure}

\begin{figure}[htb]
\vspace*{-2cm}
\hspace*{-3cm}
\includegraphics[scale=1.0]{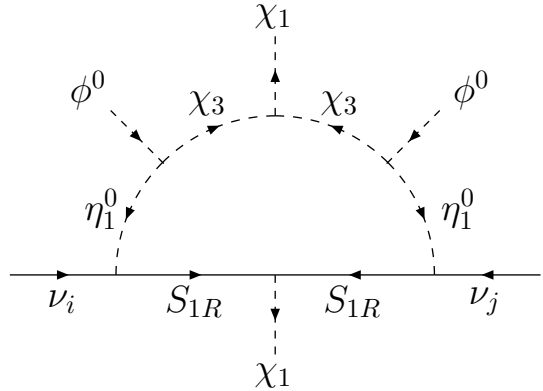}
\vspace*{-21.0cm}
\caption{One-loop active neutrino mass from $S$.}
\label{fig3}
\end{figure}

\begin{figure}[htb]
\vspace*{0cm}
\hspace*{-3cm}
\includegraphics[scale=1.0]{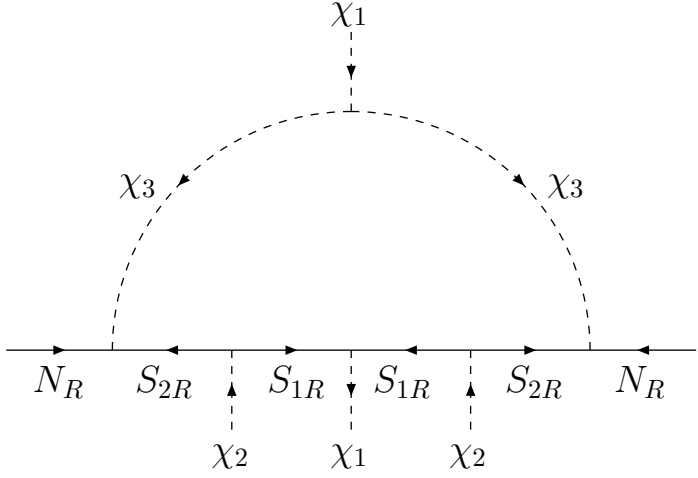}
\vspace*{-21.5cm}
\caption{One-loop sterile neutrino mass from $S$.}
\label{fig4}
\end{figure}

The two heavy fermion triplets obtain masses from the 
$\Sigma_R \Sigma_R \bar{\chi}_2^0$ interactions, whereas $S_{1R,2R}$ do 
so through $S_{1R} S_{2R} \bar{\chi}_2^0$ and $S_{1R} S_{1R} \bar{\chi}_1^0$. 
The quarks get tree-level masses from $\bar{u}_R (u_L \phi^0 - 
d_L \phi^+)$ and $(\bar{u}_L \phi^+ + \bar{d}_L \phi^0)d_R$.  Note that 
$\Phi$ is the only scalar doublet with even $Z_2$, corresponding to the 
one Higgs doublet of the SM, solely responsible for 
electroweak symmetry breaking.
The three active neutrinos $\nu_L$ and the one singlet ``sterile'' neutrino 
$N_{R}$ are massless at tree level.  They acquire radiative masses in 
one loop as shown in Figs.~\ref{fig2} to \ref{fig4}. 
The requisite couplings are  
$\bar{\Sigma}^0_R \nu_L \eta_2^0$, $\Phi^\dagger \eta_2 \bar{\chi}_3^0$, 
$\chi_3 \chi_3 \bar{\chi}_1^0$, $\bar{S}_{1R} \nu_L \eta_1^0$, 
$\Phi^\dagger \eta_1 \chi_3^0$, and $N_{R} S_{2R} \chi_3^0$.

Looking at the one-loop diagrams of Figs.~\ref{fig2} and \ref{fig3}, we 
see that instead 
of just one extra scalar doublet $\eta$ in the original scotogenic 
model~\cite{m06}, we now have two: one to couple to the two $\Sigma$'s, 
the other to $S_{1R}$.  More importantly, because of the $U(1)_X$ 
assignments of $\Sigma_R$ and $S_R$ which come from the anomaly-free 
conditions, they are odd under the unbroken residual $Z_2$ allowing the 
existence of dark matter.
At the same time, the neutral singlet fermion $N_R$ is even under $Z_2$ 
and massless at tree level, so it is suitable as a light sterile neutrino 
once it acquires a radiative mass through $S$.  Thus this anomaly-free 
$U(1)_X$ model naturally accommodates three active neutrinos and one 
sterile neutrino, all of which obtain radiative masses.  Note that the 
quartic couplings $\Phi^\dagger \eta_2 \bar{\chi}_1^0 \chi_3^0$ and 
$\Phi^\dagger \eta_1 \bar{\chi}_3^0 \chi_1^0$ are allowed, which also 
contribute to Figs.~2 and 3 respectively.

To evaluate the one-loop diagrams of Figs.~1 to 4, we note first that each is 
a sum of simple diagrams with one internal fermion line and one internal 
scalar line.  Each contribution is infinite, but the sum is finite. 
In Fig.~1, it is given by~\cite{m06}
\begin{equation}
({\cal M}_\nu)_{ij} = \sum_k {h_{ik} h_{jk} M_k \over 16 \pi^2} 
[F(m_R^2/M_k^2) - F(m_I^2/M_k^2)],
\end{equation}
where $M_k~(k=1,2,3)$ are the three $N_R$ Majorana masses, $m_R$ is 
the $\sqrt{2} Re(\eta^0)$ mass, $m_I$ is the $\sqrt{2} Im(\eta^0)$ mass, 
and $F(x) = x \ln x/(x-1)$.  
In Figs.~2 to 4, we need to consider the more complicated scalar and fermion 
sectors.  There are 8 real scalar fields, spanning 
$\sqrt{2} Re(\eta^0_{1,2})$, $\sqrt{2} Im(\eta^0_{1,2})$, $\sqrt{2} 
Re(\chi^0_3)$, $\sqrt{2} Im(\chi^0_3)$, $\sqrt{2} Re(\xi^0)$, $\sqrt{2} 
Im(\xi^0)$.  Let their mass eigenstates be $\zeta_l$ with mass $m_l$.  
There are 4 Majorana fermion fields, spanning $\Sigma^0_{1R}$, 
$\Sigma^0_{2R}$, $S_{1R}$, $S_{2R}$. Let their mass eigenstates be 
$\psi_k$ with mass $M_k$.

In Fig.~2, let the $\bar{\Sigma}^0_{1R} \nu_i 
\eta_2^0$ and $\bar{\Sigma}^0_{2R} \nu_i \eta_2^0$ couplings be $h_{i1}^{(2)}$ 
and $h_{i2}^{(2)}$, then its contribution to ${\cal M}_\nu$ is given by
\begin{eqnarray}
({\cal M}_\nu)^{(2)}_{ij} &=& {h_{i1}^{(2)} h_{j1}^{(2)} \over 16 \pi^2} 
\sum_k (z^\Sigma_{1k})^2 M_k \sum_l [(y^R_{2l})^2 F(x_{lk}) - 
(y^I_{2l})^2 F(x_{lk})] 
\nonumber \\ 
&+& {h_{i2}^{(2)} h_{j2}^{(2)} \over 16 \pi^2} \sum_k 
(z^\Sigma_{2k})^2 M_k \sum_l [(y^R_{2l})^2 F(x_{lk}) - (y^I_{2l})^2 F(x_{lk})],
\end{eqnarray}
where $\Sigma^0_{1R} = \sum_k z^\Sigma_{1k} \psi_k$, 
$\Sigma^0_{2R} = \sum_k z^\Sigma_{2k} \psi_k$, 
$\sqrt{2} Re(\eta_2^0) = \sum_l y^R_{2l} \zeta_l$, 
$\sqrt{2} Im(\eta_2^0) = \sum_l y^I_{2l} \zeta_l$, 
with $\sum_k (z^\Sigma_{1k})^2 = \sum_k (z^\Sigma_{2K})^2 = \sum_l 
(y^R_{2l})^2 = \sum_l (y^I_{2l})^2 = 1$, and $x_{lk} = m_l^2/M_k^2$.
In Fig.~3, let the $\bar{S}_{1R} \nu_i \eta_1^0$ coupling be $h_{i1}^{(1)}$, 
then its contribution to ${\cal M}_\nu$ is given by
\begin{equation}
({\cal M}_\nu)^{(1)}_{ij} = {h_{i1}^{(1)} h_{j1}^{(1)} \over 16 \pi^2} \sum_k 
(z^S_{1k})^2 M_k \sum_l [(y^R_{1l})^2 F(x_{lk}) - (y^I_{1l})^2 F(x_{lk})],
\end{equation}
where $S_{1R} = \sum_k z^S_{1k} \psi_k$,  
$\sqrt{2} Re(\eta_1^0) = \sum_l y^R_{1l} \zeta_l$, 
$\sqrt{2} Im(\eta_1^0) = \sum_l y^I_{1l} \zeta_l$, with 
$\sum_k (z^S_{1k})^2 = \sum_l (y^R_{1l})^2 = \sum_l (y^I_{1l})^2 = 1$. 
In Fig.~4, let the $S_{2R} N_R \chi_3^0$ coupling be $h_2^{(3)}$, then 
\begin{equation}
m_N = {h_{2}^{(3)} h_{2}^{(3)} \over 16 \pi^2} \sum_k 
(z^S_{2k})^2 M_k \sum_l [(y^R_{3l})^2 F(x_{lk}) - (y^I_{3l})^2 F(x_{lk})],
\end{equation}
where $S_{2R} = \sum_k z^S_{2k} \psi_k$,  
$\sqrt{2} Re(\chi_3^0) = \sum_l y^R_{3l} \zeta_l$, 
$\sqrt{2} Im(\chi_3^0) = \sum_l y^I_{3l} \zeta_l$, with 
$\sum_k (z^S_{2k})^2 = \sum_l (y^R_{3l})^2 = \sum_l (y^I_{3l})^2 = 1$.  

In the above, the three active neutrinos $\nu_{1,2,3}$ acquire masses 
through their couplings to three dark neutral fermions, i.e. $\Sigma^0_{1R}$, 
$\Sigma^0_{2R}$, $S_{1R}$, whereas the one sterile neutrino $N$ acquires 
mass through its coupling to $S_{2R}$.  However, since $S_{1R}$ mixes with 
$S_{2R}$ at tree level, there is also mixing between $\nu_i$ and $N$ as 
shown in Fig.~5, with
\begin{figure}[htb]
\vspace*{-3cm}
\hspace*{-3cm}
\includegraphics[scale=1.0]{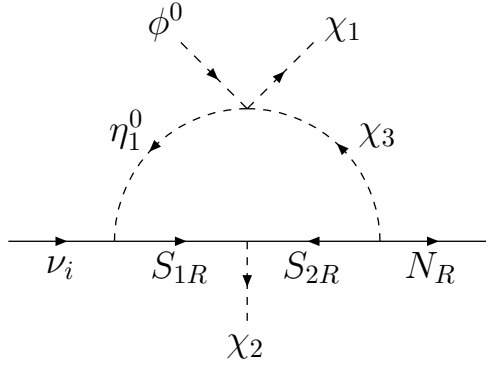}
\vspace*{-20.5cm}
\caption{One-loop active-sterile neutrino mixing from $S$.}
\label{fig5}
\end{figure}
\begin{equation}
m_{\nu N} = {h_{i1}^{(1)} (h_2^{(3)})^* \over 16 \pi^2} \sum_k z^S_{1k} 
z^S_{2k} M_k \sum_l [y_{1L}^R y_{3l}^R F(x_{lk}) - y^I_{1l} y^I_{3l} F(x_{lk}],
\end{equation}
where $\sum_k z^S_{1k} z^S_{2k} = \sum _l y^R_{1l} y^R_{3l} = \sum_l y^I_{1l} 
y^I_{3l} = 0$.   Note that the structures of these one-loop formulas are 
all similar, and there is enough freedom in choosing the various parameters 
to obtain masses of order 0.1 eV for $\nu$ and 1 eV for $N$, as well as a 
sizeable $\nu-N$ mixing.   Note also that the last term in each case 
corresponds to the cancellation among several scalars which allow the 
loops to be finite and should be naturally small.  In Fig.~1, it is 
represented by the well-known $(\lambda_5/2)(\Phi^\dagger \eta)^2 + H.c.$ 
term which splits $Re(\eta^0)$ and $Im(\eta^0)$ in mass.   
In our case for example, in Eqs.~(4) 
and (5), let $h \sim 10^{-1}$, the $\sum z^2 M$ factor $\sim 1$ TeV, 
the $\sum [(y^R)^2 - (y^I)^2] F$ factor $\sim 10^{-9}$ (which means that 
the $\bar{\chi}_1 \chi_3^2$ coupling is very small), then $m_\nu 
\sim 0.1$ eV.  In Eq.~(6), let the $\sum [(y^R)^2 - (y^I)^2] F$ factor 
$\sim 10^{-8}$ instead, then $m_N \sim 1$ eV.  In Eq.~(7), let the 
$\sum z_1 z_2 M$ factor be 100 GeV, and the $\sum [y^Ry^R - y^Iy^I] F$ factor 
$\sim 10^{-9}$ (which means that the $\eta_1^\dagger \Phi \bar{\chi}_1 \chi_3$ 
coupling is very small), then $m_{\nu N} \sim 0.1$ eV.
This is thus a possible framework for 
accommodating three active neutrinos plus a fourth light sterile neutrino, 
in the 3+1 scheme~\cite{sterileglobal}, with best fit values  
$\Delta m^2_{41} = 0.93 \; \text{eV}^2, \; \lvert U_{e4} \rvert = 0.15, \; 
\lvert U_{\mu 4} \rvert  = 0.17$, albeit having a large $\chi^2$, to ease the 
long-standing tension between $\nu_e$ appearance and $\nu_\mu$ 
disappearance experiments in the $\Delta m^2 \sim$ few eV$^2$ range.

Our model is also an example of the recently proposed 
framework~\cite{m14,m15}, where charged leptons also acquire radiative 
masses.  Indeed, they do so here also through the same four dark fermions, 
i.e. $\Sigma^0_{1R}$, $\Sigma^0_{2R}$, $S_{1R}$, $S_{2R}$, with the addition 
of a scalar triplet $\xi^{(++,+,0)}$ and a scalar singlet $\chi_4^+$, as shown 
in Figs.~6 and 7.
\begin{figure}[htb]
\vspace*{-3cm}
\hspace*{-3cm}
\includegraphics[scale=1.0]{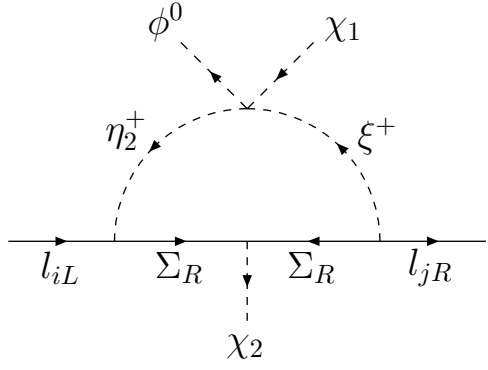}
\vspace*{-20.5cm}
\caption{One-loop charged-lepton mass from $\Sigma$.}
\label{fig6}
\end{figure}
\begin{figure}[htb]
\vspace*{-3cm}
\hspace*{-3cm}
\includegraphics[scale=1.0]{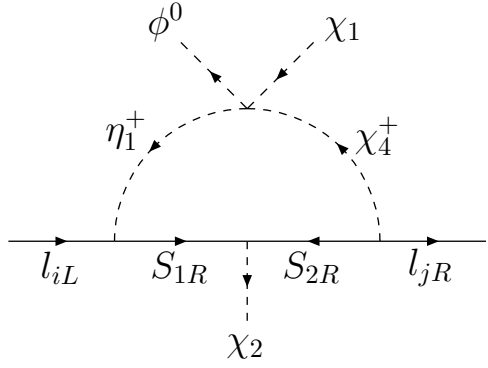}
\vspace*{-20.5cm}
\caption{One-loop charged-lepton mass from $S$.}
\label{fig7}
\end{figure}
There are 4 charged scalars, spanning $\eta_1^+$, $\eta_2^+$, $\xi^+$, 
$\chi_4^+$.  Let their mass eigenstates be $\omega_r^+$ with mass $m_r$. 
In Fig.~6, let the $l_{jR} \Sigma_{1R} \xi^+$ and $l_{jR} \Sigma_{2R} \xi^+$ 
couplings be $h^\xi_{j1}$ and $h^\xi_{j2}$, then its contribution to 
${\cal M}_l$ is
\begin{eqnarray}
({\cal M}_l)^{(\xi)}_{ij} &=& {h_{i1}^{(2)} (h^\xi_{j1})^* \over 16 \pi^2} 
\sum_k (z^\Sigma_{1k})^2 M_k \sum_r y^+_{2r} y_{\xi r} F(x_{rk}) \nonumber \\
&+& {h_{i2}^{(2)} (h^\xi_{j2})^* \over 16 \pi^2} 
\sum_k (z^\Sigma_{2k})^2 M_k \sum_r y^+_{2r} y_{\xi r} F(x_{rk}),
\end{eqnarray}
where $\eta_2^+ = \sum_r y^+_{2r} \omega^+_r$, $\xi^+ = \sum_r y_{\xi r} 
\omega_r^+$.  In Fig.~7, let the $l_{jR} S_{2R} \chi_4^+$ coupling be $h^\chi_{j}$, 
then its contribution to 
${\cal M}_l$ is
\begin{eqnarray}
({\cal M}_l)^{(\chi)}_{ij} &=& {h_{i1}^{(2)} (h^\chi_{j})^* \over 16 \pi^2} 
\sum_k z^S_{1k} z^S_{2k} M_k \sum_r y^+_{1r} y_{\chi r} F(x_{rk}), 
\end{eqnarray}
where $\eta_1^+ = \sum_r y^+_{1r} \omega^+_r$, $\chi_4^+ = \sum_r y_{\chi r} 
\omega_r^+$.  Let the $\sum y^+ y F$ factor $\sim 1$, and vary $h^{\xi,\chi}$ 
from 1 to 0.1 to 0.001, then $m_\tau,m_\mu,m_e$ may be 
obtained.  Here we require the $\eta_2 \Phi \bar{\chi}_1 \xi^\dagger$ and 
$\eta_1 \Phi \bar{\chi}_1 \chi_4$ couplings be large.  One immediate 
consequence of radiative charged-lepton masses 
is the possible significant deviation of the Higgs Yukawa coupling to 
$\bar{l} l$ from the value $m_l/$(246 GeV) required by the SM.  Detailed 
analyses~\cite{fm14,fmz15} have been performed for some specific models.

The neutral dark scalars $\zeta_l$ have in general components which are 
not electroweak singlets $(\eta^0_{1,2}, \xi^0)$.  As such, they are not 
good dark-matter candidates because their interactions with the $Z$ gauge 
boson would result in too large a cross section for their direct detection 
in underground experiments.  Hence one of the neutral dark fermions $\psi_k$ 
is a much better DM candidate.  Note that whereas $\Sigma^0_{1R}$ and 
$\Sigma^0_{2R}$ are components of $SU(2)_L$ triplets, they do not couple 
to $Z$ because they have $I_3 = 0$.  Note also that they mix with 
$S_{1R}$ and $S_{2R}$ only in one loop.  The case of $\Sigma^0$ as dark 
matter in the triplet fermion analog of the scotogenic model was discussed 
in Ref.~\cite{ms09}.  Here the important 
change is that $\Sigma^0_{1R}$ and $\Sigma^0_{2R}$ have both $SU(2)_L$ and 
$U(1)_X$ interactions.   On the other hand, suppose the lighter linear 
combination of $S_{1R}$ and $S_{2R}$ is dark matter, call it $\psi_0$, 
then only $U(1)_X$ is involved.   As shown recently in \cite{bd15, bda15}, 
the allowed region of parameter space from dark matter relic abundance, 
direct detection and collider constraints corresponds to the s-wave 
resonance region near $m_X \approx 2 m_{\psi_0}$.  
The $U(1)_X$ gauge boson mass $m_X$ in our model comes from $\langle 
\chi_{1,2} \rangle = u_{1,2}$.  If $n_1=n_4=1$ is chosen in Table 1, then 
the SM Higgs does not transform under $U(1)_X$ and there is no $X-Z$ mixing. 
In that case,
\begin{equation}
m_X^2 = 2 g_X^2 (u_1^2 + 9 u_2^2).
\end{equation}
The $U(1)_X$ charges of $(u,d)_L$, $u_R$, $d_R$, $(\nu,l)_L$ are all 1, and 
those of $l_R$, $N_R$, $\chi_1$, $\chi_2$ are $-1$, $-3$, 1, 3.  These 
particles are even under the residual $Z_2$ of $U(1)_X$.  The dark sector 
consists of fermions $\Sigma_{1R}$, $\Sigma_{2R}$, $S_{1R}$, $S_{2R}$, with 
$U(1)_X$ charges 3/2, 3/2, 1/2, 5/2, as well as scalars $\eta_1$, $\eta_2$, 
$\chi_3$, $\chi_4$, $\xi$, with $U(1)_X$ charges $-1/2$, 1/2, 1/2, $-3/2$, 
$-1/2$.

Instead of having a sterile neutrino of 1 eV, it is also possible in our 
model to make it a few keV, thus rendering $N$ a warm dark-matter 
candidate.  This may require $h_2^{(3)}$ in Eq.~(6) 
to be much greater than the corresponding Yukawa couplings in Eqs.~(4) 
and (5) for the active neutrinos.  On the other hand, $\nu-N$ mixing has 
to be much more suppressed in order not to overclose the Universe or 
conflict with observed X-ray data.   According to Ref.~\cite{kevork}, 
these may be avoided if the mixing $\lvert U_{i4} \rvert$ is less than 
$10^{-4}$.  Such a small mixing also makes the keV sterile neutrino 
long-lived on cosmological time scales.  It could also provide an explanation 
to the recently observed 3.55 keV X-ray line~\cite{Xray1} 
after analysing the data taken by the XMM-Newton X-Ray telescope in the 
spectrum of 73 galaxy clusters. The same line also appears in the 
\textit{Chandra} observations of the Perseus cluster \cite{Xray2} and 
the XMM-Newton observations of the Milky Way Centre \cite{Xray3}. In the 
absence of any astrophysical interpretation of the line due to some atomic 
transitions, the origin of this X-ray line can be explained naturally by 
sterile neutrino dark matter with mass approximately 7.1 keV decaying into 
a photon and a standard model neutrino.  As reported in Ref.~\cite{Xray2}, the 
required mixing angle of sterile neutrino with active neutrino should be 
of the order of $ \sin^2{2\theta} \approx 10^{-11}-10^{-10} $
in order to give rise to the observed X-Ray line flux. For such a tiny 
mixing angle, Fig.~5 must be strongly suppressed, implying thus almost 
zero $\chi_3 - \eta_1$ mixing.  This in turn will make Fig.~3 vanish, 
thus predicting one nearly massless active neutrino.

We have shown in this paper how a stabilizing $Z_2$ symmetry for dark matter 
may be derived from a new anomaly-free $U(1)_X$ extension of the standard 
model.   Using just the one Higgs doublet of the SM, we have also shown 
how three charged leptons and active neutrinos plus a sterile neutrino 
acquire radiative masses through the dark sector.  This explains why the 
sterile neutrino mass itself is also small. 
Apart from the possibilities of long lived sterile neutrino dark matter and 
cold dark matter separately as discussed above, our model is well-suited for 
the much more interesting mixed-dark-matter scenario, i.e. the coexistence 
of both.  Such a scenario could be important from the point of view of 
large structure formation, as well as offering proofs in different indirect 
detection experiments ranging from gamma rays to X-rays.  We leave such a 
complete analysis to future investigations.    

This work is supported in part 
by the U.~S.~Department of Energy under Grant No.~DE-SC0008541.

\baselineskip 18pt
\bibliographystyle{unsrt}

\end{document}